\documentclass[aps,prb,twocolumn,superscriptaddress,floatfix]{revtex4-2}

\usepackage{graphicx}    
\usepackage{amsmath}     
\usepackage{amssymb}     
\usepackage{hyperref}    
\usepackage{xcolor}      
\usepackage{bm}
\usepackage{array}

\begin{document}

\title{Twisting the pseudo-magnetic field lattice in periodically strained Graphene}

\author{L-M. Benneka}
\affiliation{Univ. Grenoble Alpes, CEA, Grenoble INP, IRIG, PHELIQS, 38000 Grenoble, France}

\author{T. Rhouma}
\affiliation{Laboratoire de Physique Théorique et Modélisation (UMR 8089), CY Cergy Paris Université, CNRS, Cergy-Pontoise, France}

\author{C. B. Winkelmann}
\affiliation{Univ. Grenoble Alpes, CEA, Grenoble INP, IRIG, PHELIQS, 38000 Grenoble, France}

\author{M. Zelsmann}
\affiliation{Univ. Grenoble Alpes, CNRS, CEA/LETI-Minatec, Grenoble INP, LTM, Grenoble, France}

\author{G. Trambly de Laissardière}
\affiliation{Laboratoire de Physique Théorique et Modélisation (UMR 8089), CY Cergy Paris Université, CNRS, Cergy-Pontoise, France}

\author{V.T. Renard}
\email{vincent.renard@cea.fr}
\affiliation{Univ. Grenoble Alpes, CEA, Grenoble INP, IRIG, PHELIQS, 38000 Grenoble, France}

\date{\today}

\begin{abstract}
Periodic deformations flatten electronic bands of graphene which could host strongly correlated electrons similar to those observed in twisted graphene bilayers. The most spectacular effects of strong correlations like superconductivity or magnetism have however not yet been observed despite several experimental attempts. Here, we use tight-binding calculations to investigate the effect of twisting the periodic strain lattice with respect to the atomic lattice. We find that misaligning the two lattices kills the band flattening. We also report on the universal behavior of the band flattening by periodic strain which is controlled by a single parameter, a pseudo-magnetic field flux. Our study strongly suggests that controlling the twist of the strain lattice is essential to unlock the potential of periodic strains in graphene. 
\end{abstract}

\maketitle

\section{Introduction}

Twisted graphene layers have renewed the study of strongly correlated phases as the moiré pattern provides a knob to tune the band structure of graphene all the way down to non-dispersing bands in which Coulomb interactions are the dominant energy scale \cite{Trambly2010, Bistritzer2011}. This has led to the discovery of a wealth of new interesting electronic phases \cite{Bernevig2024}. 
However, experimentally, twisted graphene bilayers are difficult to fabricate and suffer from strains inherent to the fabrication of the samples \cite{Huder2018,Mesple2021}. It is therefore interesting to develop new ways to realize strongly correlated electron physics in graphene. 

Periodic deformation of the layer has been suggested as a viable alternative to flatten graphene bands \cite{Milovanovich2020,Mao2020}. The band flattening is now well understood as resulting from the motion of electrons in the periodic pseudo-magnetic field induced by the strain patterns and could result in interesting new strongly correlated phases \cite{Uchoa2013,Chen2015,Milovanovich2020,Mao2020,Peltonen2020, Giambastiani2022, Phong2022,Mahmud2023,Wan2023,Yuan2024,Yuan2025}. The role of the symmetry of the strain pattern in decoupling the lower energy flatbands from the higher bands has been highlighted \cite{Mahmud2023} and atomic relaxation has also been considered \cite{Yuan2024} so that the non-interacting physics underlying the bands in this system is now well described.

Interactions are anticipated to lead to competing strongly correlated ground states \cite{Manesco2021}, gate tunable magnetic properties \cite{Manesco2021b} or Chern insulating states \cite{Gao2023}. 
Even under moderate repulsive interactions, the system could host an antiferromagnetic insulating ground state at half-filling. Upon doping into the flat band, the magnetic order is anticipated to lead to the stabilization of chiral singlet $d+id$ superconducting phase alongside robust ferromagnetism \cite{Peltonen2020,Meng2024}. \par
Yet, although very narrow bands should be possible \cite{Milovanovich2020} and an initial vibrant experimental demonstration of the band flattening in spontaneously buckled graphene \cite{Mao2020} and several other attempts \cite{Mi2015,Mason2018,Mason2018b,Xu2018,Rakic2024,Mason2025}, the most pronounced correlation effects like superconductivity or magnetism have still not been observed in this system. \par
In this paper, we show theoretically that band flattening in periodically strainned graphene is extremely sensitive to the relative twist between the atomic lattice and the strain lattice. This has been overlooked in previous theoretical and experimental studies and therefore appears as a significant potential explanation of the lack of strong correlation effects in experiments.

\begin{figure*}[t]
    \includegraphics[width=1\textwidth]{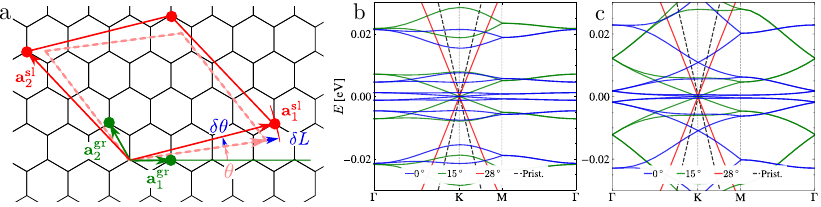}
    \caption{(a) Schematic of a rotated supercell superimposed the graphene lattice. The dashed pink cell indicates the targeted supercell of period $L$ and twist $\theta$. The solid red supercell indicates the commensurate approximation introducing errors $\delta L$ and $\delta \theta$ discussed in the text.   Electronic band structure along the path $\Gamma \mathrm{KM}\Gamma$ for different twist angles $0^\circ$ (blue), $15^\circ$ (green) and $28^\circ$ (red) at $L=20$ nm, $h_0 = 0.9$ nm and (b) $\varphi = 0^\circ$ and (c) $\varphi = 45^\circ$. Note that we present the bands at $\theta=28^\circ$ to ease the comparison between the small and large twist since the $K$ points fold at $\Gamma$ for $\theta=30^\circ$.}
    \label{fig1}
\end{figure*}
\section{Model}
\subsection{Tight-binding model}

We calculate the electronic bands of periodically strained graphene with a tight-binding model as done previously \cite{Uchoa2013,Chen2015,Milovanovich2020,Mao2020,Peltonen2020, Giambastiani2022, Phong2022,Wan2023,Yuan2024,Yuan2025, Manesco2021,Manesco2021b,Gao2023}. Hopping between nearest-neighbor graphene $\pi$ orbitals is accounted for by the following Hamiltonian \cite{Trambly2010}:

\begin{equation}
	H = \sum_{i,j} t_{ij}\:c_i^{\dagger} c_j
\end{equation}

where $t_{ij}$ is the hopping parameter between sites $i$ and $j$, $c^{(\dagger)}_i$ is the annihilation (creation) operator on site $i$.
The spatial variations of the hopping parameter resulting from the deformations are taken into account by the following dependence on the distance $d_{ij}$ between site $i$ and $j$ \cite{Trambly2010} :
\begin{equation}
	t_{ij} = t_0 \: \mathrm{exp} \left[ -\beta \left(\frac{d_{ij}}{a_{cc}}-1 \right) \right]
\end{equation}

where $t_0=-2.7$ eV is the pristine graphene hopping parameter,  $\beta\approx 3.37$ is the Grüneisen parameter and $a_{cc}=0.14$ nm is the length of unstrained carbon-carbon bonds.

Under periodic strain, the position of site $i$ can be expressed as  $\mathbf{p}_i = \mathbf{r}_i + \mathbf{u}$ where $\mathbf{u}=(u_x,u_y,h)$ is the displacement vector field evaluated at the site position $\mathbf{r}_i$ of the unstrained honeycomb lattice. In this article, periodic strain is introduced as a triangular lattice of out-of-plane displacements reading

\begin{equation}
	h(\mathbf{r})=h_0 \sum_l \cos(\mathbf{G}_l^\mathrm{sl} \cdot \mathbf{r} + \varphi)
    \label{eq : profile}
\end{equation}

where the $\mathbf{G}_l^\mathrm{sl}$ are the triangular superlattice reciprocal vectors  and $\varphi$ is a phase allowing to break $C_{2z}$ symmetry of the strain lattice for $\varphi \neq 0^\circ$ \cite{Mao2020,Milovanovich2020,Mahmud2023}. 
We used the expressions of the in-plane displacements $(u_x(h),u_y(h))$ derived in Ref.~\onlinecite{Yuan2024} by minimizing the elastic energy of the membrane in a continuum model.
The Hamiltonian described above was diagonalized consistently using Kwant package \cite{Groth2014} and a custom-developed code \cite{Trambly2010}.

\begin{figure*}[t]
    \includegraphics[width=0.96\textwidth]{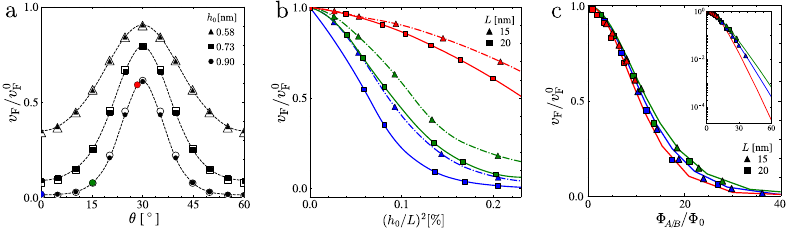}
    \caption{(a) Fermi velocity as function of the lattice orientation $\theta$ for $L=20$ nm. Filled markers correspond to $\varphi=45^\circ$ and opened markers correspond to $\varphi=0^\circ$. Dashed lines are a guide for the eyes and the colored points are the Fermi velocities corresponding to the band structures displayed in Fig.~\ref{fig1}. (b) Fermi velocity at $\theta = 0^\circ$ (blue), $\theta = 15^\circ$ (green) and $\theta = 28^\circ$ (red) calculated as function of the strain amplitude and for two different superlattice periodicities $L$. (c) Fermi velocity as a function of the flux $\Phi_{A/B}$ in half of the unit cell for different superlattice size and rotations. Symbols correspond to results from tight-binding and lines are calculated from Eq.~\ref{eq:phong}.}
    \label{fig2}
\end{figure*}

\subsection{Twisting the superlattice}

The strain lattice is superimposed on the graphene atomic lattice (Red rhombus in Fig.~\ref{fig1}a). The real-space strain-superlattice vectors $\mathbf{a}_{1,2}^\mathrm{sl}$ are usually chosen to be aligned with graphene lattice vectors ($\mathbf{a}_{1}^\mathrm{gr}=\sqrt{3}a_{cc}  (1,0) ,~\mathbf{a}_{2}^\mathrm{gr}=\sqrt{3}a_{cc}  (-1/2,\sqrt{3}/2))$ so that it is straightforward to obtain a commensurate supercell $\mathbf{a}_{i}^\mathrm{sl}= n \mathbf{a}_{i}^\mathrm{gr}$ necessary for a tight-binding calculation \cite{Milovanovich2020}. We refer to this situation as the aligned case and we now describe the procedure used to obtain commensurate supercells when the superlattice is rotated by the twist angle $\theta$ with respect to the graphene lattice.

A suitable $C_{3}$ symmetric commensurate superlattice requires two integers $(n,m)$ such that :

\begin{equation}
	\begin{bmatrix}
		\mathbf{a}_1^\mathrm{sl} \\ \mathbf{a}_2^\mathrm{sl}
	\end{bmatrix}
	=
	\begin{pmatrix}
		n & m \\ -m & n-m
	\end{pmatrix}
	\begin{bmatrix}
		\mathbf{a}^\mathrm{gr}_1 \\ \mathbf{a}^\mathrm{gr}_2
	\end{bmatrix}
\end{equation}

from which the superlattice period $L$ is expressed as :
\begin{align}
    L &= a_{cc}\sqrt{3}\sqrt{n^2+m^2-nm}
\end{align}

and the twist angle $\theta$ between the superlattice and the graphene lattice verifies :

\begin{align}
    \begin{split}
	    \cos(\theta) &= \frac{\mathbf{a}^\mathrm{gr}_1\cdot \mathbf{a}_1^\mathrm{sl} }{\rVert\mathbf{a}_1^\mathrm{gr}\rVert \cdot \lVert \mathbf{a}^\mathrm{sl}_{1} \rVert} = \frac{a_{cc}\sqrt{3}}{L}\left(n - \frac{m}{2} \right)\\
        \sin(\theta) &= \frac{\mathbf{a}_1^\mathrm{gr,\perp} \cdot\mathbf{a}^\mathrm{sl}_1}{\lVert\mathbf{a}_1^\mathrm{gr,\perp}\rVert \cdot \lVert \mathbf{a}^\mathrm{sl}_{1} \rVert} = \frac{3a_{cc}}{2L}m 
    \end{split}
    \label{eq : Theta commensurate}
\end{align}
where $\mathbf{a}_{1}^\mathrm{gr,\perp}=\sqrt{3}a_{cc}(0,1)$. 
Inverting Eqs.~\ref{eq : Theta commensurate} allows to find the values of $(n,m)$ for a desired superlattice period and twist:

\begin{equation}
	\begin{aligned}
		n &= \frac{L}{\sqrt{3} a_{\mathrm{cc}}} \left( \mathrm{cos}(\theta) + \frac{\mathrm{sin}(\theta)}{\sqrt{3}} \right)  \\
		m &=  \frac{2L}{3 a_{\mathrm{cc}}} \mathrm{sin}(\theta)
	\end{aligned}
	\label{eqTr2}
\end{equation}

The numbers $n$ and $m$ are not integers for all targeted values of $L$ and $\theta$ and we achieve commensurability by rounding them to the closest integer. This corresponds to choosing the nearest lattice site (Fig.~\ref{fig1}a) which introduces an error in the length of the superlattice. This error is of the order of $\delta L/L \approx a_{cc}/L$ which is no more than 5\% for superlattices larger than 5 nm and is therefore neglected. The procedure also discretizes the available values of $\theta$, which occurence in the interval $[0, \pi/6]$ is $\mathcal{N}^\theta = \frac{2\pi}{9}\frac{L}{a_{cc}}$, sufficiently large for $L\gg a_{cc}$. 

\section{Results}
\subsection{Effect of the twist}

Figures~\ref{fig1}b and c show the main result of our study: the evolution of the band structure of periodically strained graphene with the relative alignment of the atomic and strain lattice. We plot the band structures obtained for $L=20$ nm and $h_0=0.9$ nm for $\varphi=0$ and $\varphi=45^\circ$. While the bands flatten for all values of twist, it is obvious that the effect is much more pronounced when the two lattices are aligned. The effect of the phase $\varphi$ is to isolate the lowest energy bands from the higher energy bands owing to the breaking of $C_{2z}$ symmetry \cite{Mahmud2023} but the band flattening at the Dirac point is independent of $\varphi$. 

In order to perform a systematic study, we follow previous works on twisted graphene bilayers in using the Fermi velocity $v_F$ measured at the Dirac point to quantify the flatness of the bands \cite{Bistritzer2011}. Figure~\ref{fig2}a shows $v_F/v_F^0$ as a function of $\theta$ for various corrugation amplitudes and fixed $L=20$ nm ($v_F^{0}$ is the Fermi velocity of pristine graphene). The Fermi velocity is $60^\circ$ periodic as expected from the symmetry of the lattices. Consistently with Fig.~\ref{fig1}b-c, $v_F$ exhibits a minimum for zigzag alignment ($\theta=0^\circ$) and a maximum  for armchair alignment ($\theta=30^\circ$) where the bands are almost immune to strain. The maximum in $v_F$ is sharper than its minima so that band flattening is effective in an interval of $\pm10^\circ$. Figure~\ref{fig2}a shows that the band flattening at the Dirac point does not depend on the phase $\varphi$ whose influence is therefore only to decouple the low energy bands from the upper ones \cite{Mahmud2023}.

Figure~\ref{fig2}b shows that, at a given twist angle, the evolution of band flattening with twist and strain is consistent with previous results~ \cite{Milovanovich2020}: the larger the strain and the larger the lattice period, the flatter the bands.

\subsection{Scaling of the renormalization of \texorpdfstring{$v_F$}{TEXT}}

\begin{figure*}[t]
    \includegraphics[width=1.01\textwidth]{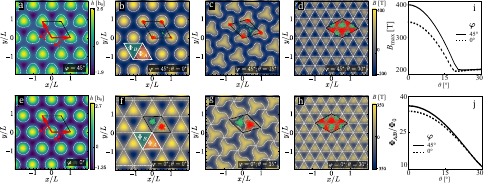}
    \caption{(a) $h$-displacement field for $h_0=0.9$ nm, $L=20$ nm,  $\theta=0^\circ$ for $\varphi=45^\circ$. The same is plotted for  $\varphi=0^\circ$ in (e). The superlattice unit cell is highlighted in black. The corresponding pseudo-magnetic fields are shown in (b) and (f) with white lines showing $B=$0 T contours. The unit cell used for the calculation of the magnetic flux is shown in white. The field pattern for $\theta=15^\circ$ and $\theta=30^\circ$ are shown in (c,g) and (d,h).  The low-energy density of states is shown in  red (green) for lattices A (B). (i) Evolution of the maximum of pseudo-magnetic field  with $\theta$. (j) Same for the magnetic flux $\Phi_{A/B}$.}
    \label{fig3}
\end{figure*}

One may wonder why twisting the strain lattice with respect to the atomic lattice has such a strong effect on the band flattening. We will now show that it is possible to understand this behavior and also the strain and lattice size dependence with a single parameter: a pseudo-magnetic field flux.

At low strains ($<$10\%), variations of  hopping imposed by strain act on Dirac electrons as a local pseudo-magnetic field arising from a pseudo-vector potential $\mathbf{A}$, which is directly connected to the strain tensor $\epsilon$ by \cite{Neto2009,Guinea22010,Amorim2016}:
\begin{equation}
	\mathbf{A}
	=-\frac{\beta}{2ea_{cc}} 
	\begin{pmatrix}
		\epsilon_{xx} - \epsilon_{yy} \\ -2\epsilon_{xy}
	\end{pmatrix}.
	\label{eqTrA}
\end{equation}

It appears in the low-energy Hamiltonian $H = \hbar v_F \boldsymbol{\sigma} \cdot(\mathbf{q} + e\mathbf{A})$. Including in-plane relaxations, \cite{Yuan2024}, the vertical displacement profile of Eq.~\ref{eq : profile} leads to the pseudo-magnetic field $\mathbf{B}(\mathbf{r})=\mathbf{\mathbf{\nabla}\times}\mathbf{A}(\mathbf{r})$ with the following norm at $\theta=0^\circ$:
\begin{equation}
    B(\mathbf{r}) = B_0 \sum_l \sin(\mathbf{G}_l^\mathrm{sl} \cdot \mathbf{r} + 2\varphi)
    \label{eq:b}
\end{equation}

where $B_0 = -\frac{4\pi^3 \beta (1+\nu) \hbar}{\sqrt{3}e a_{cc}}\frac{h_0^2}{L^3}$ with $\nu$ the Poisson coefficient. At $\theta\neq 0$ one has to evaluate it numerically.  

The pseudo-magnetic flux is used to quantify the effect of strain \cite{Gao2023,Giambastiani2022}. It is usually estimated using $B_0$ or $B_{max}$ multiplied by the surface of the unit cell because, owing to the conservation of time reversal symmetry, which is not broken by strain, the flux on the unit cell is strictly 0. We argue that the effect of strain is better described by the flux on half of the unit cell where the pseudo-magnetic field is predominantly of a given sign (See Fig.~\ref{fig3}). This is further justified by the fact that the wavefunctions are localized on one sublattice for a given sign of the pseudo-magnetic field.\cite{Schneider2015,Mao2020} Hence sublattice-polarized states effectively occupy half of the unit cell and we label the corresponding fluxes $\Phi_{A/B}$. This is illustrated in Fig~\ref{fig3} where the sublattice-resolved density of states is plotted on top of the pseudo-magnetic field landscape.

With this definition, Fig.~\ref{fig2}c shows that all the curves of Figs.~\ref{fig2}a and \ref{fig2}b collapse on a single one when plotted as a function of the ratio of $\Phi_{A/B}$ and the flux quantum $\Phi_0=h/e$.  The increase of $v_F$ with increasing $\theta$ then suggests that the pseudo-magnetic field flux decreases with $\theta$. Figure~\ref{fig3} shows that the pattern of the pseudo-magnetic fields evolves strongly with the twist of the strain lattice. Figure~\ref{fig3}i shows that the amplitude of the pseudo-magnetic field decreases between $\theta=0$ and $\theta\approx20^\circ$. This is the main driving force of the reduction of the flux in this regime. The pseudo-magnetic field saturates above $\theta\approx20^\circ$ but yet the pseudo-magnetic flux continues to decrease steadily (Fig.~\ref{fig3}j). In this regime, the flux reduction is due to a reshaping of the pseudo-magnetic landscape which eventually evolves towards a doubling of its periodicity at $\theta=30^\circ$ ($L\rightarrow L/\sqrt{3} \approx 0.577L$). In each half unit cell, the flux originating from the pseudo-magnetic field of dominant sign is gradually compensated by an increasing contribution of pseudo-magnetic field of the other sign delocalizing electrons. 

The universal behavior of Fig.~\ref{fig2}c can be recovered from an analytical calculation derived from a Aharonov-Casher development \cite{Aharonov1979,Phong2025,Sourrouille2025} for a periodic magnetic field for which the velocity is renormalized as : 
\begin{equation}
    \frac{v_F}{v_F^0}=\frac{|S|}{\mathcal{N^+}\mathcal{N^-}  }.
\label{eq:phong}
\end{equation}

Here $\mathcal{N}^\pm = [\int_S exp(\pm 2e\tilde{\Phi}(\bm{r})/\hbar)d^2\bm{r}]^{1/2}$ are the Bloch zero-mode normalization amplitudes and the local flux $\tilde{\Phi}(\bm{r})=-\sum_{G}\tilde{B}_{\mathbf{G}}e^{i\mathbf{G}\cdot\mathbf{r}}/|G|^2$. Applying this formula to our case of pseudo-magnetic field describes very well our numerical calculations as illustrated by the colored lines in Fig.~\ref{fig2}c. At $\theta=0^\circ$, the asymptotic behavior at large flux can be established from Laplace's method for integral approximation \cite{Wong2001} shown in appendix. It is expressed as $\mathcal{N^\pm}=[\int_S d^2\mathbf{r}\:\exp(\pm \lambda\:g(\mathbf{r}))]^{1/2}$ with $\lambda=\Phi/\Phi_0=4\pi(B_0^\theta/G^2)/\Phi_0$ and $g(\mathbf{r})=\sum_l \sin(\mathbf{G}_l\cdot\mathbf{r} + 2\varphi)$.
\begin{equation}
    \frac{v_F}{v_F^0} \sim N \sqrt{\frac{\Phi}{\Phi_0}}\exp\left(-M\frac{\Phi}{\Phi_0}\right) \:\:\: \text{as }\:\:\Phi \gg \Phi_0
    \label{eq:asy}
\end{equation}
with $N\approx 0.20$ and $M=2.25-2.59$ slightly depending on the spatial phase $\varphi$. Hence, the log-scaled velocity renormalization is mostly linear in the large flux limit. Equation~\ref{eq:phong} implies that the bands are never completely flat since the Dirac cones remain, which is confirmed by the asymptotic behavior. Yet, the bands can be flat enough to access the strong Coulomb interaction regime \cite{Milovanovich2020,Gao2023}. 
At $\theta\neq0^\circ$ a numerical estimation of Eq.~\ref{eq:phong} shows that the twist has additional small non-monotonous contribution to the velocity renormalization as observed Fig.~\ref{fig2}c in log scale. Further investigations are needed to understand fully this effect.

\section{Discussion}
\subsection{Effect of the strain-induced scalar potential}
Until now, we have neglected that strain also creates a scalar potential in addition to the vector potential \cite{CastroNeto2009}. This contribution enters the Hamiltonian of the system as a periodic on-site energy reading:
\begin{equation}
	V_{s}(\mathbf{r})=g_s\cdot(\varepsilon_{xx}(\mathbf{r})+\varepsilon_{yy}(\mathbf{r}))
    \label{eq : scalar}
\end{equation}
with $g_s$ a constant of few eV according to recent experiments \cite{Wang2021} and theoretical estimates \cite{Grassano2020}
.

\begin{figure}[t]
    \includegraphics[width=0.9\columnwidth]{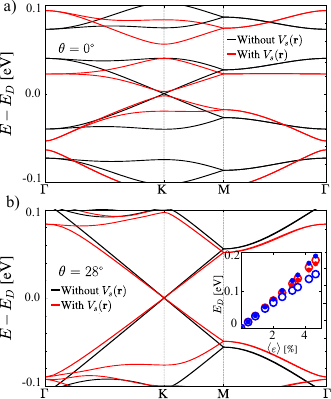}
    \caption{Comparison of the band structure with (red lines) and without (black lines) the deformation potential $V_s(\mathbf{r})$ for $g_s=4$ eV, $L=20$ nm, $h_0 = 0.51$ nm, $\varphi=45^\circ$ and (a) $\theta = 0^\circ$ and (b) $\theta = 28^\circ$. The bands calculated including the scalar potential were shifted vertically to facilitate the comparison. The inset shows the shift of the Dirac Point with the average strain $\langle \varepsilon \rangle$ inside the unit cell for $\theta = 0^\circ$ in blue and $\theta=30^\circ$ in red. The filled markers correspond to $\varphi=45^\circ$ and opened markers correspond to $\varphi=0^\circ$.}
    \label{fig4}
\end{figure}

Our calculations show that the main effect of the scalar potential is to shift rigidly the energy bands by an amount proportional to the average biaxial strain needed to expand the graphene surface to the corrugated surface (inset to Fig.~\ref{fig4}). Figure~\ref{fig4} shows the bands after removal of this shift to allow direct comparison with the bands obtained without the potential. At large twist, the scalar potential has a very small effect. Its influence is larger in the aligned case where it opens a gap at the Dirac point thanks to the $C_{2z}$ symmetry breaking lifting the cones chirality. This is consistent with the behavior predicted for an external fluctuating potential \cite{Gao2023} and deserves a detailed study on its own. All in all, Fig.~\ref{fig4} confirms that the alignment of the atomic and strain lattice leads to the narrowest bands.

\subsection{Twist of the strain lattice in previous experiments}
The results presented above indicate that the strain lattice orientation is a crucial parameter for the band flattening in periodically strained graphene. Fourier  transforming the STM images of Ref.~\cite{Mao2020} reveal that the strain lattice was indeed aligned with the crystallographic directions. We believe that this arises from the fact that the buckling was induced by a biaxial strain applied to graphene and that it was naturally relaxed along the crystallographic directions. This explains why the effects of periodic strain were particularly pronounced in this study. As for the other experimental studies of periodically strained graphene \cite{Mason2018,HSU2020,Rakic2024,Kammarchedu2024}, we were not able to determine the strain lattice orientation but it could very well be that it is part of the cause of the weak effects of strain observed. 

\subsection{Design of future experiments}
The universal curve of Fig.~\ref{fig2} demonstrates that similar Fermi velocities -and therefore flatbands- can be achieved independently of the size of the strain lattice and its orientation, if sufficiently large strain is applied. This could be used to design experiments. However one should also keep in mind that not everything can be done in practice and that the strain also has to comply with the mechanical limits of the graphene material but also with the fact that experimentally, the graphene membrane will conform to the substrate only if the elastic energy remains smaller than the adhesion energy \cite{Reserbat-Plantey2014}.

Finally, the lattice rotation method and the scaling parameter introduced in this work are readily extendable to the investigation of other one-dimensional (1D) and two-dimensional (2D) strain patterns and material systems. The primary challenge arises from the commensurability condition between both lattices. However, the application of a small uniaxial strain can restore the boundary conditions in cases where the lattice symmetries are mutually incompatible. \par

\section*{Acknowledgements}
This work was supported by the ANR Flatmoi project (ANR-21-CE30-0029) and by Agence Nationale de la Recherche support under the France 2030 heading, bearing the reference ANR-24-RRII-0004 in the framework of high-risk research program Audace! led by the CEA.

\appendix*

\section{Asymptotic behavior with Laplace's method}

As developed in the Letter, we would be interested in the explicit formulation of the Fermi velocity \ref{eq:phong}, i.e., we need to find the expression of the integral $\mathcal{N}^\pm = [\int_S exp(\pm 2e\Phi(\bm{r})/\hbar)d^2\bm{r}]^{1/2}$. The problem can be reduced to the formulation $\mathcal{I}^\pm = \int_S exp(\pm \lambda\:g(\mathbf{r}))d^2\mathbf{r}$ where $S$ is a bounded surface, $\lambda>0$ and $g(\mathbf{r})$ is at least $C^2(\mathbb{R}^n)$ and bounded.

For instance, if $g(\mathbf{r}) =g_0 \sum_l \sin(\mathbf{G}_l \cdot \mathbf{r} + 2\varphi)$, $\mathcal{I}^\pm$ becomes an integral which can be expressed from the introduction of the Jacobi–Anger expansion. It expresses a link between $z \in \mathbb{C} \mapsto e^{i z \cos(\theta)}$ and the series of Bessel functions of the first kind $I_n(z)$. 
\begin{equation*}
    e^{i z \sin\theta}
    = \sum_{n} I_n(z)\, e^{i n \theta}
\end{equation*}

The integral can be fully expressed as 
\begin{equation*}
\begin{split}
    \mathcal{I}^\pm
    & = \sum_{n,m,p} I_n(i \lambda g_0)\:I_m(i\lambda g_0)\:I_p(i\lambda g_0)\, \exp(2i(n+m+p)\phi) \\ 
    & \quad \cdot\int_S dS \exp(i(n\mathbf{G}^\text{sl}_1 + m\mathbf{G}^\text{sl}_2 + p\mathbf{G}^\text{sl}_3)\cdot \mathbf{r})
\end{split}
\end{equation*}
The Bessel functions and non-orthogonality of $\mathbf{G}_{l=1,2,3}$ prevent the integrals from separating and find a simple expression of $\mathcal{I}^\pm$. However, an analytical relation exists by formulating its asymptotic behavior for $\lambda \gg 1$.
We assume that $g(\mathbf{r})$ admits a unique maximum since it is bounded over the surface $S\backslash \partial S$ and we name its position $\mathbf{r}_0$. The Taylor expansion around $\mathbf{r}_0$ produces the relation $g(\mathbf{r}) \approx g(\mathbf{r}_0) - \frac{1}{2} (\mathbf{r}-\mathbf{r}_0)^T|\nabla^2g|(\mathbf{r}_0)\:(\mathbf{r}-\mathbf{r}_0)$.

We introduce the Hessian matrix $|\nabla^2g|$ as $n\times n$ positive definite and symmetric which expresses the local concavity around the maximum value of $g(\mathbf{r})$. Laplace's method appears by identifying a Gaussian integral with a width proportional to $1/\lambda$ as exponential of the maximum value of $g(\mathbf{r})$. It acts as the main contribution in the integral. Hence integrated over the surface $S$, we can approximate the integral of the Gaussian function over all $\mathbb{R}^n$ for large enough $\lambda$. 
\begin{equation*}
\begin{split}
    \mathcal{I}^\pm &\sim e^{\lambda g(\mathbf{r_0})} \int_{\mathbb{R}^n} \exp\left( - \frac{\lambda}{2} (\mathbf{r}-\mathbf{r_0})^T |\nabla^2 g|(\mathbf{r}_0) (\mathbf{r}-\mathbf{r}_0) \right) \\
    &\sim e^{\lambda g(\mathbf{r}_0)}  \frac{(2\pi/\lambda)^{n/2}}{\sqrt{\det(|\nabla^2 g|(\mathbf{r}_0))}}
\end{split}
\end{equation*}
The last step introduces the number $p_\pm$, the equivalent maximum (resp. minimum) of $\pm g(\mathbf{r})$. Also, this method works for positive or negative $\lambda$ since the sign can be absorbed in $g$ and in the Hessian matrix of the maximum (resp. minimum) of $g$. Eventually the integral of interest becomes
\begin{equation*}
    \mathcal{I}^\pm \sim \left[\frac{2\pi\:p_\pm^2}{\det(|\nabla^2 (\pm g)|)} \frac{e^{2\lambda \|\pm g\|_\infty}}{\lambda} \right]^{1/4}
\end{equation*}
And the Fermi velocity eventually from Eq.\ref{eq:phong} reads
\begin{equation*}
\begin{split}
    \frac{v_F}{v_F^0} & \sim N \sqrt{\lambda} \: e^{-M \lambda } \\
    N & = S \left[ \frac{\det(|\nabla^2 g|)\cdot\det(|\nabla^2 (\mathrm{-}g)|)}{(2\pi\:p_+ p_-)^2}\right]^{1/4}\\
    M & = \frac{\| g\|_\infty +  \| \mathrm{-}g\|_\infty}{2}
\end{split}
\end{equation*}

\vspace{12pt} 
\noindent\rule{\linewidth}{0.0pt}
\vspace{12pt} 

\bibliographystyle{apsrev4-2}
\bibliography{Biblio_psg}

\begin{thebibliography}{41}%
\makeatletter
\providecommand \@ifxundefined [1]{%
 \@ifx{#1\undefined}
}%
\providecommand \@ifnum [1]{%
 \ifnum #1\expandafter \@firstoftwo
 \else \expandafter \@secondoftwo
 \fi
}%
\providecommand \@ifx [1]{%
 \ifx #1\expandafter \@firstoftwo
 \else \expandafter \@secondoftwo
 \fi
}%
\providecommand \natexlab [1]{#1}%
\providecommand \enquote  [1]{``#1''}%
\providecommand \bibnamefont  [1]{#1}%
\providecommand \bibfnamefont [1]{#1}%
\providecommand \citenamefont [1]{#1}%
\providecommand \href@noop [0]{\@secondoftwo}%
\providecommand \href [0]{\begingroup \@sanitize@url \@href}%
\providecommand \@href[1]{\@@startlink{#1}\@@href}%
\providecommand \@@href[1]{\endgroup#1\@@endlink}%
\providecommand \@sanitize@url [0]{\catcode `\\12\catcode `\$12\catcode
  `\&12\catcode `\#12\catcode `\^12\catcode `\_12\catcode `\%12\relax}%
\providecommand \@@startlink[1]{}%
\providecommand \@@endlink[0]{}%
\providecommand \url  [0]{\begingroup\@sanitize@url \@url }%
\providecommand \@url [1]{\endgroup\@href {#1}{\urlprefix }}%
\providecommand \urlprefix  [0]{URL }%
\providecommand \Eprint [0]{\href }%
\providecommand \doibase [0]{https://doi.org/}%
\providecommand \selectlanguage [0]{\@gobble}%
\providecommand \bibinfo  [0]{\@secondoftwo}%
\providecommand \bibfield  [0]{\@secondoftwo}%
\providecommand \translation [1]{[#1]}%
\providecommand \BibitemOpen [0]{}%
\providecommand \bibitemStop [0]{}%
\providecommand \bibitemNoStop [0]{.\EOS\space}%
\providecommand \EOS [0]{\spacefactor3000\relax}%
\providecommand \BibitemShut  [1]{\csname bibitem#1\endcsname}%
\let\auto@bib@innerbib\@empty
\bibitem [{\citenamefont {Trambly~de Laissardière}\ \emph
  {et~al.}(2010)\citenamefont {Trambly~de Laissardière}, \citenamefont
  {Mayou},\ and\ \citenamefont {Magaud}}]{Trambly2010}%
  \BibitemOpen
  \bibfield  {author} {\bibinfo {author} {\bibfnamefont {G.}~\bibnamefont
  {Trambly~de Laissardière}}, \bibinfo {author} {\bibfnamefont
  {D.}~\bibnamefont {Mayou}},\ and\ \bibinfo {author} {\bibfnamefont
  {L.}~\bibnamefont {Magaud}},\ }\href {https://doi.org/10.1021/nl902948m}
  {\bibfield  {journal} {\bibinfo  {journal} {Nano Letters}\ }\textbf {\bibinfo
  {volume} {10}},\ \bibinfo {pages} {804} (\bibinfo {year} {2010})}\BibitemShut
  {NoStop}%
\bibitem [{\citenamefont {Bistritzer}\ and\ \citenamefont
  {MacDonald}(2011)}]{Bistritzer2011}%
  \BibitemOpen
  \bibfield  {author} {\bibinfo {author} {\bibfnamefont {R.}~\bibnamefont
  {Bistritzer}}\ and\ \bibinfo {author} {\bibfnamefont {A.~H.}\ \bibnamefont
  {MacDonald}},\ }\href {https://www.pnas.org/doi/full/10.1073/pnas.1108174108}
  {\bibfield  {journal} {\bibinfo  {journal} {PNAS}\ }\textbf {\bibinfo
  {volume} {108}},\ \bibinfo {pages} {12233 12237} (\bibinfo {year}
  {2011})}\BibitemShut {NoStop}%
\bibitem [{\citenamefont {Bernevig}\ and\ \citenamefont
  {Efetov}(2024)}]{Bernevig2024}%
  \BibitemOpen
  \bibfield  {author} {\bibinfo {author} {\bibfnamefont {B.~A.}\ \bibnamefont
  {Bernevig}}\ and\ \bibinfo {author} {\bibfnamefont {D.~K.}\ \bibnamefont
  {Efetov}},\ }\href {https://doi.org/10.1063/pt.jvsd.yhyd} {\bibfield
  {journal} {\bibinfo  {journal} {Physics Today}\ }\textbf {\bibinfo {volume}
  {77}},\ \bibinfo {pages} {38} (\bibinfo {year} {2024})}\BibitemShut {NoStop}%
\bibitem [{\citenamefont {Huder}\ \emph {et~al.}(2018)\citenamefont {Huder},
  \citenamefont {Artaud}, \citenamefont {Quang}, \citenamefont
  {de~Laissardi\`ere}, \citenamefont {Chapelier},\ and\ \citenamefont
  {Renard}}]{Huder2018}%
  \BibitemOpen
  \bibfield  {author} {\bibinfo {author} {\bibfnamefont {L.}~\bibnamefont
  {Huder}}, \bibinfo {author} {\bibfnamefont {A.}~\bibnamefont {Artaud}},
  \bibinfo {author} {\bibfnamefont {T.~L.}\ \bibnamefont {Quang}}, \bibinfo
  {author} {\bibfnamefont {G.~T.}\ \bibnamefont {de~Laissardi\`ere}}, \bibinfo
  {author} {\bibfnamefont {C.}~\bibnamefont {Chapelier}},\ and\ \bibinfo
  {author} {\bibfnamefont {V.~T.}\ \bibnamefont {Renard}},\ }\href
  {https://doi.org/10.1103/PhysRevLett.120.156405} {\bibfield  {journal}
  {\bibinfo  {journal} {Physical Review Letters}\ }\textbf {\bibinfo {volume}
  {120}},\ \bibinfo {pages} {156405} (\bibinfo {year} {2018})}\BibitemShut
  {NoStop}%
\bibitem [{\citenamefont {Mesple}\ \emph {et~al.}(2021)\citenamefont {Mesple},
  \citenamefont {Missaoui}, \citenamefont {Cea}, \citenamefont {Huder},
  \citenamefont {Guinea}, \citenamefont {Trambly~de Laissardi\`ere},
  \citenamefont {Chapelier},\ and\ \citenamefont {Renard}}]{Mesple2021}%
  \BibitemOpen
  \bibfield  {author} {\bibinfo {author} {\bibfnamefont {F.}~\bibnamefont
  {Mesple}}, \bibinfo {author} {\bibfnamefont {A.}~\bibnamefont {Missaoui}},
  \bibinfo {author} {\bibfnamefont {T.}~\bibnamefont {Cea}}, \bibinfo {author}
  {\bibfnamefont {L.}~\bibnamefont {Huder}}, \bibinfo {author} {\bibfnamefont
  {F.}~\bibnamefont {Guinea}}, \bibinfo {author} {\bibfnamefont
  {G.}~\bibnamefont {Trambly~de Laissardi\`ere}}, \bibinfo {author}
  {\bibfnamefont {C.}~\bibnamefont {Chapelier}},\ and\ \bibinfo {author}
  {\bibfnamefont {V.~T.}\ \bibnamefont {Renard}},\ }\href
  {https://doi.org/10.1103/PhysRevLett.127.126405} {\bibfield  {journal}
  {\bibinfo  {journal} {Phys. Rev. Lett.}\ }\textbf {\bibinfo {volume} {127}},\
  \bibinfo {pages} {126405} (\bibinfo {year} {2021})}\BibitemShut {NoStop}%
\bibitem [{\citenamefont {Milovanovich}\ \emph {et~al.}(2020)\citenamefont
  {Milovanovich}, \citenamefont {Andelkovic}, \citenamefont {L.},\ and\
  \citenamefont {Peeters}}]{Milovanovich2020}%
  \BibitemOpen
  \bibfield  {author} {\bibinfo {author} {\bibfnamefont {S.}~\bibnamefont
  {Milovanovich}}, \bibinfo {author} {\bibfnamefont {M.}~\bibnamefont
  {Andelkovic}}, \bibinfo {author} {\bibfnamefont {C.}~\bibnamefont {L.}},\
  and\ \bibinfo {author} {\bibfnamefont {F.~M.}\ \bibnamefont {Peeters}},\
  }\href {https://doi.org/10.1103/PhysRevB.102.245427} {\bibfield  {journal}
  {\bibinfo  {journal} {Phys. Rev. B}\ }\textbf {\bibinfo {volume} {102}},\
  \bibinfo {pages} {245427} (\bibinfo {year} {2020})}\BibitemShut {NoStop}%
\bibitem [{\citenamefont {Mao}\ and\ \citenamefont {\textit{et
  al.}}(2020)}]{Mao2020}%
  \BibitemOpen
  \bibfield  {author} {\bibinfo {author} {\bibfnamefont {J.}~\bibnamefont
  {Mao}}\ and\ \bibinfo {author} {\bibnamefont {\textit{et al.}}},\ }\href
  {https://doi.org/10.1038/s41586-020-2567-3} {\bibfield  {journal} {\bibinfo
  {journal} {Nature}\ }\textbf {\bibinfo {volume} {584}},\ \bibinfo {pages}
  {215} (\bibinfo {year} {2020})}\BibitemShut {NoStop}%
\bibitem [{\citenamefont {Uchoa}\ and\ \citenamefont
  {Barlas}(2013)}]{Uchoa2013}%
  \BibitemOpen
  \bibfield  {author} {\bibinfo {author} {\bibfnamefont {B.}~\bibnamefont
  {Uchoa}}\ and\ \bibinfo {author} {\bibfnamefont {Y.}~\bibnamefont {Barlas}},\
  }\href {https://doi.org/10.1103/PhysRevLett.111.046604} {\bibfield  {journal}
  {\bibinfo  {journal} {Phys. Rev. Lett.}\ }\textbf {\bibinfo {volume} {111}},\
  \bibinfo {pages} {046604} (\bibinfo {year} {2013})}\BibitemShut {NoStop}%
\bibitem [{\citenamefont {Si}\ \emph {et~al.}(2016)\citenamefont {Si},
  \citenamefont {Sun},\ and\ \citenamefont {Liu}}]{Chen2015}%
  \BibitemOpen
  \bibfield  {author} {\bibinfo {author} {\bibfnamefont {C.}~\bibnamefont
  {Si}}, \bibinfo {author} {\bibfnamefont {Z.}~\bibnamefont {Sun}},\ and\
  \bibinfo {author} {\bibfnamefont {F.}~\bibnamefont {Liu}},\ }\href
  {https://doi.org/10.1039/C5NR07755A} {\bibfield  {journal} {\bibinfo
  {journal} {Nanoscale}\ }\textbf {\bibinfo {volume} {8}},\ \bibinfo {pages}
  {3207} (\bibinfo {year} {2016})}\BibitemShut {NoStop}%
\bibitem [{\citenamefont {Peltonen}\ and\ \citenamefont
  {Heikkilä}(2020)}]{Peltonen2020}%
  \BibitemOpen
  \bibfield  {author} {\bibinfo {author} {\bibfnamefont {T.~J.}\ \bibnamefont
  {Peltonen}}\ and\ \bibinfo {author} {\bibfnamefont {T.~T.}\ \bibnamefont
  {Heikkilä}},\ }\href {https://doi.org/10.1088/1361-648X/ab8b9d} {\bibfield
  {journal} {\bibinfo  {journal} {Journal of Physics: Condensed Matter}\
  }\textbf {\bibinfo {volume} {32}},\ \bibinfo {pages} {365603} (\bibinfo
  {year} {2020})}\BibitemShut {NoStop}%
\bibitem [{\citenamefont {V}\ \emph {et~al.}(2022)\citenamefont {V},
  \citenamefont {Colangelo}, \citenamefont {Tredicucci}, \citenamefont
  {Roddaro},\ and\ \citenamefont {Pitanti}}]{Giambastiani2022}%
  \BibitemOpen
  \bibfield  {author} {\bibinfo {author} {\bibfnamefont {D.}~\bibnamefont {V}},
  \bibinfo {author} {\bibfnamefont {F.}~\bibnamefont {Colangelo}}, \bibinfo
  {author} {\bibfnamefont {A.}~\bibnamefont {Tredicucci}}, \bibinfo {author}
  {\bibfnamefont {S.}~\bibnamefont {Roddaro}},\ and\ \bibinfo {author}
  {\bibfnamefont {A.}~\bibnamefont {Pitanti}},\ }\href
  {https://doi.org/10.1063/5.0080098} {\bibfield  {journal} {\bibinfo
  {journal} {Appl. Phys. Lett}\ }\textbf {\bibinfo {volume} {131}},\ \bibinfo
  {pages} {085103} (\bibinfo {year} {2022})}\BibitemShut {NoStop}%
\bibitem [{\citenamefont {Phong}\ and\ \citenamefont {Mele}(2022)}]{Phong2022}%
  \BibitemOpen
  \bibfield  {author} {\bibinfo {author} {\bibfnamefont {V.~T.}\ \bibnamefont
  {Phong}}\ and\ \bibinfo {author} {\bibfnamefont {E.~J.}\ \bibnamefont
  {Mele}},\ }\href {https://doi.org/10.1103/PhysRevLett.128.176406} {\bibfield
  {journal} {\bibinfo  {journal} {Phys. Rev. Lett.}\ }\textbf {\bibinfo
  {volume} {128}},\ \bibinfo {pages} {176406} (\bibinfo {year}
  {2022})}\BibitemShut {NoStop}%
\bibitem [{\citenamefont {Mahmud}\ \emph {et~al.}(2023)\citenamefont {Mahmud},
  \citenamefont {Zhai},\ and\ \citenamefont {Sandler}}]{Mahmud2023}%
  \BibitemOpen
  \bibfield  {author} {\bibinfo {author} {\bibfnamefont {M.~T.}\ \bibnamefont
  {Mahmud}}, \bibinfo {author} {\bibfnamefont {D.}~\bibnamefont {Zhai}},\ and\
  \bibinfo {author} {\bibfnamefont {N.}~\bibnamefont {Sandler}},\ }\href
  {https://doi.org/10.1021/acs.nanolett.3c02513} {\bibfield  {journal}
  {\bibinfo  {journal} {Nanoletters}\ }\textbf {\bibinfo {volume} {23}},\
  \bibinfo {pages} {7725} (\bibinfo {year} {2023})}\BibitemShut {NoStop}%
\bibitem [{\citenamefont {Wan}\ \emph {et~al.}(2023)\citenamefont {Wan},
  \citenamefont {Sarkar}, \citenamefont {Sun},\ and\ \citenamefont
  {Lin}}]{Wan2023}%
  \BibitemOpen
  \bibfield  {author} {\bibinfo {author} {\bibfnamefont {X.}~\bibnamefont
  {Wan}}, \bibinfo {author} {\bibfnamefont {S.}~\bibnamefont {Sarkar}},
  \bibinfo {author} {\bibfnamefont {K.}~\bibnamefont {Sun}},\ and\ \bibinfo
  {author} {\bibfnamefont {S.-Z.}\ \bibnamefont {Lin}},\ }\href
  {https://doi.org/10.1103/PhysRevB.108.125129} {\bibfield  {journal} {\bibinfo
   {journal} {Phys. Rev. B}\ }\textbf {\bibinfo {volume} {108}},\ \bibinfo
  {pages} {125129} (\bibinfo {year} {2023})}\BibitemShut {NoStop}%
\bibitem [{\citenamefont {Yuan}\ and\ \citenamefont {Zhu}(2024)}]{Yuan2024}%
  \BibitemOpen
  \bibfield  {author} {\bibinfo {author} {\bibfnamefont {X.}~\bibnamefont
  {Yuan}}\ and\ \bibinfo {author} {\bibfnamefont {S.}~\bibnamefont {Zhu}},\
  }\href {https://doi.org/10.1103/PhysRevB.109.245408} {\bibfield  {journal}
  {\bibinfo  {journal} {Phys. Rev. B}\ }\textbf {\bibinfo {volume} {109}},\
  \bibinfo {pages} {245408} (\bibinfo {year} {2024})}\BibitemShut {NoStop}%
\bibitem [{\citenamefont {Yuan}\ \emph {et~al.}(2025)\citenamefont {Yuan},
  \citenamefont {Zhang},\ and\ \citenamefont {Zhu}}]{Yuan2025}%
  \BibitemOpen
  \bibfield  {author} {\bibinfo {author} {\bibfnamefont {X.}~\bibnamefont
  {Yuan}}, \bibinfo {author} {\bibfnamefont {Z.}~\bibnamefont {Zhang}},\ and\
  \bibinfo {author} {\bibfnamefont {S.}~\bibnamefont {Zhu}},\ }\href
  {https://doi.org/10.1103/PhysRevB.111.235407} {\bibfield  {journal} {\bibinfo
   {journal} {Phys. Rev. B}\ }\textbf {\bibinfo {volume} {111}},\ \bibinfo
  {pages} {235407} (\bibinfo {year} {2025})}\BibitemShut {NoStop}%
\bibitem [{\citenamefont {Manesco}\ and\ \citenamefont
  {Lado}(2021)}]{Manesco2021}%
  \BibitemOpen
  \bibfield  {author} {\bibinfo {author} {\bibfnamefont {A.~L.~R.}\
  \bibnamefont {Manesco}}\ and\ \bibinfo {author} {\bibfnamefont {J.~L.}\
  \bibnamefont {Lado}},\ }\href {https://doi.org/10.1088/2053-1583/ac0b48}
  {\bibfield  {journal} {\bibinfo  {journal} {2D Materials}\ }\textbf {\bibinfo
  {volume} {8}},\ \bibinfo {pages} {035057} (\bibinfo {year}
  {2021})}\BibitemShut {NoStop}%
\bibitem [{\citenamefont {Manesco}\ \emph {et~al.}(2020)\citenamefont
  {Manesco}, \citenamefont {Lado}, \citenamefont {Ribeiro}, \citenamefont
  {Weber},\ and\ \citenamefont {Rodrigues~Jr}}]{Manesco2021b}%
  \BibitemOpen
  \bibfield  {author} {\bibinfo {author} {\bibfnamefont {A.~L.~R.}\
  \bibnamefont {Manesco}}, \bibinfo {author} {\bibfnamefont {J.~L.}\
  \bibnamefont {Lado}}, \bibinfo {author} {\bibfnamefont {E.~V.~S.}\
  \bibnamefont {Ribeiro}}, \bibinfo {author} {\bibfnamefont {G.}~\bibnamefont
  {Weber}},\ and\ \bibinfo {author} {\bibfnamefont {D.}~\bibnamefont
  {Rodrigues~Jr}},\ }\href {https://doi.org/10.1088/2053-1583/abbc5f}
  {\bibfield  {journal} {\bibinfo  {journal} {2D Materials}\ }\textbf {\bibinfo
  {volume} {8}},\ \bibinfo {pages} {015011} (\bibinfo {year}
  {2020})}\BibitemShut {NoStop}%
\bibitem [{\citenamefont {Gao}\ \emph {et~al.}(2023)\citenamefont {Gao},
  \citenamefont {Dong}, \citenamefont {Ledwith}, \citenamefont {Parker},\ and\
  \citenamefont {Khalaf}}]{Gao2023}%
  \BibitemOpen
  \bibfield  {author} {\bibinfo {author} {\bibfnamefont {Q.}~\bibnamefont
  {Gao}}, \bibinfo {author} {\bibfnamefont {J.}~\bibnamefont {Dong}}, \bibinfo
  {author} {\bibfnamefont {P.}~\bibnamefont {Ledwith}}, \bibinfo {author}
  {\bibfnamefont {D.}~\bibnamefont {Parker}},\ and\ \bibinfo {author}
  {\bibfnamefont {E.}~\bibnamefont {Khalaf}},\ }\href
  {https://doi.org/10.1103/PhysRevLett.131.096401} {\bibfield  {journal}
  {\bibinfo  {journal} {Phys. Rev. Lett.}\ }\textbf {\bibinfo {volume} {131}},\
  \bibinfo {pages} {096401} (\bibinfo {year} {2023})}\BibitemShut {NoStop}%
\bibitem [{\citenamefont {Meng}\ \emph {et~al.}(2024)\citenamefont {Meng},
  \citenamefont {Ma}, \citenamefont {Ma},\ and\ \citenamefont
  {Lin}}]{Meng2024}%
  \BibitemOpen
  \bibfield  {author} {\bibinfo {author} {\bibfnamefont {J.}~\bibnamefont
  {Meng}}, \bibinfo {author} {\bibfnamefont {R.}~\bibnamefont {Ma}}, \bibinfo
  {author} {\bibfnamefont {T.}~\bibnamefont {Ma}},\ and\ \bibinfo {author}
  {\bibfnamefont {H.-Q.}\ \bibnamefont {Lin}},\ }\bibfield  {journal} {\bibinfo
   {journal} {Physical Review B}\ }\textbf {\bibinfo {volume} {110}},\ \href
  {https://doi.org/10.1103/physrevb.110.235128} {10.1103/physrevb.110.235128}
  (\bibinfo {year} {2024})\BibitemShut {NoStop}%
\bibitem [{\citenamefont {Mi}\ \emph {et~al.}(2015)\citenamefont {Mi},
  \citenamefont {Mikael}, \citenamefont {Liu}, \citenamefont {Seo},
  \citenamefont {Gui}, \citenamefont {Ma}, \citenamefont {Nealey},\ and\
  \citenamefont {Ma}}]{Mi2015}%
  \BibitemOpen
  \bibfield  {author} {\bibinfo {author} {\bibfnamefont {H.}~\bibnamefont
  {Mi}}, \bibinfo {author} {\bibfnamefont {S.}~\bibnamefont {Mikael}}, \bibinfo
  {author} {\bibfnamefont {C.-C.}\ \bibnamefont {Liu}}, \bibinfo {author}
  {\bibfnamefont {J.-H.}\ \bibnamefont {Seo}}, \bibinfo {author} {\bibfnamefont
  {G.}~\bibnamefont {Gui}}, \bibinfo {author} {\bibfnamefont {A.~L.}\
  \bibnamefont {Ma}}, \bibinfo {author} {\bibfnamefont {P.~F.}\ \bibnamefont
  {Nealey}},\ and\ \bibinfo {author} {\bibfnamefont {Z.}~\bibnamefont {Ma}},\
  }\href {https://doi.org/10.1063/1.4932657} {\bibfield  {journal} {\bibinfo
  {journal} {Applied Physics Letters}\ }\textbf {\bibinfo {volume} {107}},\
  \bibinfo {pages} {143107} (\bibinfo {year} {2015})}\BibitemShut {NoStop}%
\bibitem [{\citenamefont {Zhang}\ \emph
  {et~al.}(2018{\natexlab{a}})\citenamefont {Zhang}, \citenamefont {Kim},
  \citenamefont {Gilbert},\ and\ \citenamefont {Mason}}]{Mason2018}%
  \BibitemOpen
  \bibfield  {author} {\bibinfo {author} {\bibfnamefont {Y.}~\bibnamefont
  {Zhang}}, \bibinfo {author} {\bibfnamefont {Y.}~\bibnamefont {Kim}}, \bibinfo
  {author} {\bibfnamefont {M.~J.}\ \bibnamefont {Gilbert}},\ and\ \bibinfo
  {author} {\bibfnamefont {N.}~\bibnamefont {Mason}},\ }\href
  {https://doi.org/10.1038/s41699-018-0076-0} {\bibfield  {journal} {\bibinfo
  {journal} {npj 2D Materials and Applications}\ }\textbf {\bibinfo {volume}
  {2}},\ \bibinfo {pages} {31} (\bibinfo {year}
  {2018}{\natexlab{a}})}\BibitemShut {NoStop}%
\bibitem [{\citenamefont {Zhang}\ \emph
  {et~al.}(2018{\natexlab{b}})\citenamefont {Zhang}, \citenamefont {Heiranian},
  \citenamefont {Janicek}, \citenamefont {Budrikis}, \citenamefont {Zapperi},
  \citenamefont {Huang}, \citenamefont {Johnson}, \citenamefont {Aluru},
  \citenamefont {Lyding},\ and\ \citenamefont {Mason}}]{Mason2018b}%
  \BibitemOpen
  \bibfield  {author} {\bibinfo {author} {\bibfnamefont {Y.}~\bibnamefont
  {Zhang}}, \bibinfo {author} {\bibfnamefont {M.}~\bibnamefont {Heiranian}},
  \bibinfo {author} {\bibfnamefont {B.}~\bibnamefont {Janicek}}, \bibinfo
  {author} {\bibfnamefont {Z.}~\bibnamefont {Budrikis}}, \bibinfo {author}
  {\bibfnamefont {S.}~\bibnamefont {Zapperi}}, \bibinfo {author} {\bibfnamefont
  {P.~Y.}\ \bibnamefont {Huang}}, \bibinfo {author} {\bibfnamefont {H.~T.}\
  \bibnamefont {Johnson}}, \bibinfo {author} {\bibfnamefont {N.~R.}\
  \bibnamefont {Aluru}}, \bibinfo {author} {\bibfnamefont {J.~W.}\ \bibnamefont
  {Lyding}},\ and\ \bibinfo {author} {\bibfnamefont {N.}~\bibnamefont
  {Mason}},\ }\href {https://doi.org/10.1021/acs.nanolett.8b00273} {\bibfield
  {journal} {\bibinfo  {journal} {Nano Letters}\ }\textbf {\bibinfo {volume}
  {18}},\ \bibinfo {pages} {2098} (\bibinfo {year}
  {2018}{\natexlab{b}})}\BibitemShut {NoStop}%
\bibitem [{\citenamefont {Xu}\ and\ \citenamefont {\textit{et
  al.}}(2018)}]{Xu2018}%
  \BibitemOpen
  \bibfield  {author} {\bibinfo {author} {\bibfnamefont {T.}~\bibnamefont
  {Xu}}\ and\ \bibinfo {author} {\bibnamefont {\textit{et al.}}},\ }\href
  {https://doi.org/10.1039/c8nr00678d} {\bibfield  {journal} {\bibinfo
  {journal} {Nanoscale}\ }\textbf {\bibinfo {volume} {10}},\ \bibinfo {pages}
  {7519} (\bibinfo {year} {2018})}\BibitemShut {NoStop}%
\bibitem [{\citenamefont {Šrut Rakić}\ \emph {et~al.}(2025)\citenamefont
  {Šrut Rakić}, \citenamefont {Gilbert}, \citenamefont {Sarkar},
  \citenamefont {Aishwarya}, \citenamefont {Polini}, \citenamefont {Madhavan},\
  and\ \citenamefont {Mason}}]{Rakic2024}%
  \BibitemOpen
  \bibfield  {author} {\bibinfo {author} {\bibfnamefont {I.}~\bibnamefont
  {Šrut Rakić}}, \bibinfo {author} {\bibfnamefont {M.~J.}\ \bibnamefont
  {Gilbert}}, \bibinfo {author} {\bibfnamefont {P.}~\bibnamefont {Sarkar}},
  \bibinfo {author} {\bibfnamefont {A.}~\bibnamefont {Aishwarya}}, \bibinfo
  {author} {\bibfnamefont {M.}~\bibnamefont {Polini}}, \bibinfo {author}
  {\bibfnamefont {V.}~\bibnamefont {Madhavan}},\ and\ \bibinfo {author}
  {\bibfnamefont {N.}~\bibnamefont {Mason}},\ }\href
  {https://doi.org/10.1021/acs.nanolett.4c03542} {\bibfield  {journal}
  {\bibinfo  {journal} {Nano Letters}\ }\textbf {\bibinfo {volume} {25}},\
  \bibinfo {pages} {41} (\bibinfo {year} {2025})}\BibitemShut {NoStop}%
\bibitem [{\citenamefont {Zhang}\ \emph
  {et~al.}(2018{\natexlab{c}})\citenamefont {Zhang}, \citenamefont {Heiranian},
  \citenamefont {Janicek}, \citenamefont {Budrikis}, \citenamefont {Zapperi},
  \citenamefont {Huang}, \citenamefont {Johnson}, \citenamefont {Aluru},
  \citenamefont {Lyding},\ and\ \citenamefont {Mason}}]{Mason2025}%
  \BibitemOpen
  \bibfield  {author} {\bibinfo {author} {\bibfnamefont {Y.}~\bibnamefont
  {Zhang}}, \bibinfo {author} {\bibfnamefont {M.}~\bibnamefont {Heiranian}},
  \bibinfo {author} {\bibfnamefont {B.}~\bibnamefont {Janicek}}, \bibinfo
  {author} {\bibfnamefont {Z.}~\bibnamefont {Budrikis}}, \bibinfo {author}
  {\bibfnamefont {S.}~\bibnamefont {Zapperi}}, \bibinfo {author} {\bibfnamefont
  {P.~Y.}\ \bibnamefont {Huang}}, \bibinfo {author} {\bibfnamefont {H.~T.}\
  \bibnamefont {Johnson}}, \bibinfo {author} {\bibfnamefont {N.~R.}\
  \bibnamefont {Aluru}}, \bibinfo {author} {\bibfnamefont {J.~W.}\ \bibnamefont
  {Lyding}},\ and\ \bibinfo {author} {\bibfnamefont {N.}~\bibnamefont
  {Mason}},\ }\href {https://doi.org/10.1021/acs.nanolett.8b00273} {\bibfield
  {journal} {\bibinfo  {journal} {Nano Letters}\ }\textbf {\bibinfo {volume}
  {18}},\ \bibinfo {pages} {2098} (\bibinfo {year}
  {2018}{\natexlab{c}})}\BibitemShut {NoStop}%
\bibitem [{\citenamefont {Groth}\ \emph {et~al.}(2014)\citenamefont {Groth},
  \citenamefont {Wimmer}, \citenamefont {Akhmerov},\ and\ \citenamefont
  {Waintal}}]{Groth2014}%
  \BibitemOpen
  \bibfield  {author} {\bibinfo {author} {\bibfnamefont {C.~W.}\ \bibnamefont
  {Groth}}, \bibinfo {author} {\bibfnamefont {M.}~\bibnamefont {Wimmer}},
  \bibinfo {author} {\bibfnamefont {A.~R.}\ \bibnamefont {Akhmerov}},\ and\
  \bibinfo {author} {\bibfnamefont {X.}~\bibnamefont {Waintal}},\ }\href
  {https://doi.org/10.1088/1367-2630/16/6/063065} {\bibfield  {journal}
  {\bibinfo  {journal} {New Journal of Physics}\ }\textbf {\bibinfo {volume}
  {16}},\ \bibinfo {pages} {063065} (\bibinfo {year} {2014})}\BibitemShut
  {NoStop}%
\bibitem [{\citenamefont {Castro~Neto}\ \emph
  {et~al.}(2009{\natexlab{a}})\citenamefont {Castro~Neto}, \citenamefont
  {Guinea}, \citenamefont {Peres}, \citenamefont {Novoselov},\ and\
  \citenamefont {Geim}}]{Neto2009}%
  \BibitemOpen
  \bibfield  {author} {\bibinfo {author} {\bibfnamefont {A.~H.}\ \bibnamefont
  {Castro~Neto}}, \bibinfo {author} {\bibfnamefont {F.}~\bibnamefont {Guinea}},
  \bibinfo {author} {\bibfnamefont {N.~M.~R.}\ \bibnamefont {Peres}}, \bibinfo
  {author} {\bibfnamefont {K.~S.}\ \bibnamefont {Novoselov}},\ and\ \bibinfo
  {author} {\bibfnamefont {A.~K.}\ \bibnamefont {Geim}},\ }\href
  {https://doi.org/10.1103/RevModPhys.81.109} {\bibfield  {journal} {\bibinfo
  {journal} {Rev. Mod. Phys.}\ }\textbf {\bibinfo {volume} {81}},\ \bibinfo
  {pages} {109} (\bibinfo {year} {2009}{\natexlab{a}})}\BibitemShut {NoStop}%
\bibitem [{\citenamefont {Guinea}\ \emph {et~al.}(2010)\citenamefont {Guinea},
  \citenamefont {Katsnelson},\ and\ \citenamefont {Geim}}]{Guinea22010}%
  \BibitemOpen
  \bibfield  {author} {\bibinfo {author} {\bibfnamefont {F.}~\bibnamefont
  {Guinea}}, \bibinfo {author} {\bibfnamefont {M.~I.}\ \bibnamefont
  {Katsnelson}},\ and\ \bibinfo {author} {\bibfnamefont {A.~K.}\ \bibnamefont
  {Geim}},\ }\href {https://doi.org/10.1038/nphys1420} {\bibfield  {journal}
  {\bibinfo  {journal} {Nature Physics}\ }\textbf {\bibinfo {volume} {6}},\
  \bibinfo {pages} {30} (\bibinfo {year} {2010})}\BibitemShut {NoStop}%
\bibitem [{\citenamefont {Amorim}\ \emph {et~al.}(2016)\citenamefont {Amorim},
  \citenamefont {Cortijo}, \citenamefont {de~Juan}, \citenamefont {Grushin},
  \citenamefont {Guinea}, \citenamefont {Gutiérrez-Rubio}, \citenamefont
  {Ochoa}, \citenamefont {Parente}, \citenamefont {Roldán}, \citenamefont
  {San-Jose}, \citenamefont {Schiefele}, \citenamefont {Sturla},\ and\
  \citenamefont {Vozmediano}}]{Amorim2016}%
  \BibitemOpen
  \bibfield  {author} {\bibinfo {author} {\bibfnamefont {B.}~\bibnamefont
  {Amorim}}, \bibinfo {author} {\bibfnamefont {A.}~\bibnamefont {Cortijo}},
  \bibinfo {author} {\bibfnamefont {F.}~\bibnamefont {de~Juan}}, \bibinfo
  {author} {\bibfnamefont {A.}~\bibnamefont {Grushin}}, \bibinfo {author}
  {\bibfnamefont {F.}~\bibnamefont {Guinea}}, \bibinfo {author} {\bibfnamefont
  {A.}~\bibnamefont {Gutiérrez-Rubio}}, \bibinfo {author} {\bibfnamefont
  {H.}~\bibnamefont {Ochoa}}, \bibinfo {author} {\bibfnamefont
  {V.}~\bibnamefont {Parente}}, \bibinfo {author} {\bibfnamefont
  {R.}~\bibnamefont {Roldán}}, \bibinfo {author} {\bibfnamefont
  {P.}~\bibnamefont {San-Jose}}, \bibinfo {author} {\bibfnamefont
  {J.}~\bibnamefont {Schiefele}}, \bibinfo {author} {\bibfnamefont
  {M.}~\bibnamefont {Sturla}},\ and\ \bibinfo {author} {\bibfnamefont
  {M.}~\bibnamefont {Vozmediano}},\ }\href
  {https://doi.org/10.1016/j.physrep.2015.12.006} {\bibfield  {journal}
  {\bibinfo  {journal} {Physics Reports}\ }\textbf {\bibinfo {volume} {617}},\
  \bibinfo {pages} {1–54} (\bibinfo {year} {2016})}\BibitemShut {NoStop}%
\bibitem [{\citenamefont {Schneider}\ \emph {et~al.}(2015)\citenamefont
  {Schneider}, \citenamefont {Faria}, \citenamefont {Viola~Kusminskiy},\ and\
  \citenamefont {Sandler}}]{Schneider2015}%
  \BibitemOpen
  \bibfield  {author} {\bibinfo {author} {\bibfnamefont {M.}~\bibnamefont
  {Schneider}}, \bibinfo {author} {\bibfnamefont {D.}~\bibnamefont {Faria}},
  \bibinfo {author} {\bibfnamefont {S.}~\bibnamefont {Viola~Kusminskiy}},\ and\
  \bibinfo {author} {\bibfnamefont {N.}~\bibnamefont {Sandler}},\ }\href
  {https://doi.org/10.1103/PhysRevB.91.161407} {\bibfield  {journal} {\bibinfo
  {journal} {Phys. Rev. B}\ }\textbf {\bibinfo {volume} {91}},\ \bibinfo
  {pages} {161407} (\bibinfo {year} {2015})}\BibitemShut {NoStop}%
\bibitem [{\citenamefont {Aharonov}\ and\ \citenamefont
  {Casher}(1979)}]{Aharonov1979}%
  \BibitemOpen
  \bibfield  {author} {\bibinfo {author} {\bibfnamefont {Y.}~\bibnamefont
  {Aharonov}}\ and\ \bibinfo {author} {\bibfnamefont {A.}~\bibnamefont
  {Casher}},\ }\href {https://doi.org/10.1103/PhysRevA.19.2461} {\bibfield
  {journal} {\bibinfo  {journal} {Phys. Rev. A}\ }\textbf {\bibinfo {volume}
  {19}},\ \bibinfo {pages} {2461} (\bibinfo {year} {1979})}\BibitemShut
  {NoStop}%
\bibitem [{\citenamefont {Phong}\ and\ \citenamefont {Mele}(2025)}]{Phong2025}%
  \BibitemOpen
  \bibfield  {author} {\bibinfo {author} {\bibfnamefont {V.~o.~T.}\
  \bibnamefont {Phong}}\ and\ \bibinfo {author} {\bibfnamefont {E.~J.}\
  \bibnamefont {Mele}},\ }\href {https://doi.org/10.1103/PhysRevB.111.125129}
  {\bibfield  {journal} {\bibinfo  {journal} {Phys. Rev. B}\ }\textbf {\bibinfo
  {volume} {111}},\ \bibinfo {pages} {125129} (\bibinfo {year}
  {2025})}\BibitemShut {NoStop}%
\bibitem [{\citenamefont {Sourrouille}(2025)}]{Sourrouille2025}%
  \BibitemOpen
  \bibfield  {author} {\bibinfo {author} {\bibfnamefont {L.}~\bibnamefont
  {Sourrouille}},\ }\href {https://arxiv.org/abs/2509.24028} {\  (\bibinfo
  {year} {2025})},\ \Eprint {https://arxiv.org/abs/2509.24028}
  {arXiv:2509.24028 [quant-ph]} \BibitemShut {NoStop}%
\bibitem [{\citenamefont {Wong}(2001)}]{Wong2001}%
  \BibitemOpen
  \bibfield  {author} {\bibinfo {author} {\bibfnamefont {R.}~\bibnamefont
  {Wong}},\ }\bibinfo {title} {9. higher dimensional integrals},\ in\ \href
  {https://doi.org/10.1137/1.9780898719260.ch9} {\emph {\bibinfo {booktitle}
  {Asymptotic Approximations of Integrals}}}\ (\bibinfo {year} {2001})\ pp.\
  \bibinfo {pages} {477--515}\BibitemShut {NoStop}%
\bibitem [{\citenamefont {Castro~Neto}\ \emph
  {et~al.}(2009{\natexlab{b}})\citenamefont {Castro~Neto}, \citenamefont
  {Guinea}, \citenamefont {Peres}, \citenamefont {Novoselov},\ and\
  \citenamefont {Geim}}]{CastroNeto2009}%
  \BibitemOpen
  \bibfield  {author} {\bibinfo {author} {\bibfnamefont {A.~H.}\ \bibnamefont
  {Castro~Neto}}, \bibinfo {author} {\bibfnamefont {F.}~\bibnamefont {Guinea}},
  \bibinfo {author} {\bibfnamefont {N.~M.~R.}\ \bibnamefont {Peres}}, \bibinfo
  {author} {\bibfnamefont {K.~S.}\ \bibnamefont {Novoselov}},\ and\ \bibinfo
  {author} {\bibfnamefont {A.~K.}\ \bibnamefont {Geim}},\ }\href
  {https://doi.org/10.1103/RevModPhys.81.109} {\bibfield  {journal} {\bibinfo
  {journal} {Rev. Mod. Phys.}\ }\textbf {\bibinfo {volume} {81}},\ \bibinfo
  {pages} {109} (\bibinfo {year} {2009}{\natexlab{b}})}\BibitemShut {NoStop}%
\bibitem [{\citenamefont {Wang}\ and\ \citenamefont {\textit{et
  al.}}(2021)}]{Wang2021}%
  \BibitemOpen
  \bibfield  {author} {\bibinfo {author} {\bibfnamefont {L.}~\bibnamefont
  {Wang}}\ and\ \bibinfo {author} {\bibnamefont {\textit{et al.}}},\ }\href
  {https://doi.org//10.1038/s42005-021-00651-y} {\bibfield  {journal} {\bibinfo
   {journal} {Communications Physics}\ }\textbf {\bibinfo {volume} {4}},\
  \bibinfo {pages} {147} (\bibinfo {year} {2021})}\BibitemShut {NoStop}%
\bibitem [{\citenamefont {Grassano}\ \emph {et~al.}(2020)\citenamefont
  {Grassano}, \citenamefont {D'Alessandro}, \citenamefont {Pulci},
  \citenamefont {Sharapov}, \citenamefont {Gusynin},\ and\ \citenamefont
  {Varlamov}}]{Grassano2020}%
  \BibitemOpen
  \bibfield  {author} {\bibinfo {author} {\bibfnamefont {D.}~\bibnamefont
  {Grassano}}, \bibinfo {author} {\bibfnamefont {M.}~\bibnamefont
  {D'Alessandro}}, \bibinfo {author} {\bibfnamefont {O.}~\bibnamefont {Pulci}},
  \bibinfo {author} {\bibfnamefont {S.~G.}\ \bibnamefont {Sharapov}}, \bibinfo
  {author} {\bibfnamefont {V.~P.}\ \bibnamefont {Gusynin}},\ and\ \bibinfo
  {author} {\bibfnamefont {A.~A.}\ \bibnamefont {Varlamov}},\ }\href
  {https://doi.org/10.1103/PhysRevB.101.245115} {\bibfield  {journal} {\bibinfo
   {journal} {Phys. Rev. B}\ }\textbf {\bibinfo {volume} {101}},\ \bibinfo
  {pages} {245115} (\bibinfo {year} {2020})}\BibitemShut {NoStop}%
\bibitem [{\citenamefont {Hsu}\ \emph {et~al.}(2020)\citenamefont {Hsu},
  \citenamefont {Teague}, \citenamefont {Wang},\ and\ \citenamefont
  {Yeh}}]{HSU2020}%
  \BibitemOpen
  \bibfield  {author} {\bibinfo {author} {\bibfnamefont {C.-C.}\ \bibnamefont
  {Hsu}}, \bibinfo {author} {\bibfnamefont {M.~L.}\ \bibnamefont {Teague}},
  \bibinfo {author} {\bibfnamefont {J.-Q.}\ \bibnamefont {Wang}},\ and\
  \bibinfo {author} {\bibfnamefont {N.-C.}\ \bibnamefont {Yeh}},\ }\href
  {https://doi.org/10.1126/sciadv.aat9488} {\bibfield  {journal} {\bibinfo
  {journal} {Science Advances}\ }\textbf {\bibinfo {volume} {6}},\ \bibinfo
  {pages} {eaat9488} (\bibinfo {year} {2020})}\BibitemShut {NoStop}%
\bibitem [{\citenamefont {Kammarchedu}\ \emph {et~al.}(2024)\citenamefont
  {Kammarchedu}, \citenamefont {Butler}, \citenamefont {Rashid}, \citenamefont
  {Ebrahimi},\ and\ \citenamefont {Kayyalha}}]{Kammarchedu2024}%
  \BibitemOpen
  \bibfield  {author} {\bibinfo {author} {\bibfnamefont {V.}~\bibnamefont
  {Kammarchedu}}, \bibinfo {author} {\bibfnamefont {D.}~\bibnamefont {Butler}},
  \bibinfo {author} {\bibfnamefont {A.~S.}\ \bibnamefont {Rashid}}, \bibinfo
  {author} {\bibfnamefont {A.}~\bibnamefont {Ebrahimi}},\ and\ \bibinfo
  {author} {\bibfnamefont {M.}~\bibnamefont {Kayyalha}},\ }\href
  {https://doi.org/10.1088/1361-6528/ad7853} {\bibfield  {journal} {\bibinfo
  {journal} {Nanotechnology}\ }\textbf {\bibinfo {volume} {35}},\ \bibinfo
  {pages} {495701} (\bibinfo {year} {2024})}\BibitemShut {NoStop}%
\bibitem [{\citenamefont {Reserbat-Plantey}\ and\ \citenamefont {\textit{et
  al.}}(2014)}]{Reserbat-Plantey2014}%
  \BibitemOpen
  \bibfield  {author} {\bibinfo {author} {\bibfnamefont {A.}~\bibnamefont
  {Reserbat-Plantey}}\ and\ \bibinfo {author} {\bibnamefont {\textit{et
  al.}}},\ }\href {https://doi.org/10.1021/nl5016552} {\bibfield  {journal}
  {\bibinfo  {journal} {Nano Letters}\ }\textbf {\bibinfo {volume} {14}},\
  \bibinfo {pages} {5044} (\bibinfo {year} {2014})}\BibitemShut {NoStop}%
\end{thebibliography}%

\end{document}